\documentclass[]{spie}  

\def\ra{\rangle}
\def\la{\langle}
\def\bege{\begin{equation}}
\def\ende{\end{equation}}
\def\begarr{\begin{eqnarray}}
\def\endarr{\end{eqnarray}}
\def\no{\noindent}
\def\non{\nonumber}

\def\ad{{{\hat a}^\dagger}}
\def\bd{{{\hat b}^\dagger}}

\def\ha{{\hat{a}}}
\def\hb{{\hat{b}}}

\def\he{{\hat{e}}}

\def\hi{\hangindent=25pt}

\usepackage[]{graphicx,color,epsfig}

\title{
Quantum Computing, Metrology, and Imaging} 

\author{
Hwang Lee, 
Pavel Lougovski, and Jonathan P. Dowling
\skiplinehalf
Department of Physics and Astronomy \\
Hearne Institute for Theoretical Physics \\
Louisiana State University,
Baton Rouge, LA~~70803-4001
}
\authorinfo{
\hi
SPIE's International Symposium on Fluctuations and Noise,
Austin, TX (May, 2005).
}

\begin{document} 
\maketitle 


\begin{abstract}

Information science is entering into a new era in which 
certain subtleties of quantum mechanics enables large enhancements 
in computational efficiency and communication security.
Naturally, precise control of quantum systems required for 
the implementation of quantum information processing protocols
implies potential breakthoughs in other sciences and technologies.
We discuss recent developments in quantum control in optical systems
and their applications in metrology and imaging.

\end{abstract}

\keywords{
Quantum computing, quantum entanglement, 
optical interferometers, High NOON states, Heisenberg limit
}


\section{INTRODUCTION}
\label{sect:intro}  

A new concept of quantum computation has been developed 
in recent years. 
The basic unit of a quantum computer is 
a quantum mechanical two-level system (qubit) 
that can be in coherent superpositions of the logical values 0 and 1, 
as opposed to classical bits that represent either the values 0 or 1.
Moreover, qubits can possess mutually tied properties while
separated in space, so-called quantum entanglement.
The implementation of computations is carried out 
by unitary transformations, 
which consist of the individual quantum logic gates. 
The utilization of superposition and entanglement leads to 
a high degree of parallelism, 
which makes the speed of certain types of computation 
exponentially faster than the classical counterpart \cite{bouwmeester00}. 

It has been known for a while that quantum entanglement has 
the potential to revolutionize 
the entire field of interferometry by providing many orders 
of magnitude improvement in interferometer sensitivity.
In particular, without nonlocal entanglement, a generic classical 
interferometer has a statistical-sampling shot-noise limited sensitivity 
However, if carefully prepared quantum correlations are engineered 
between the particles, 
then the interferometer sensitivity improves beyond the shot-noise limit.

A few years ago we introduced the Quantum Rosetta Stone 
by addressing the formal equivalence between a generic quantum logic gate,
a Ramsey-type atomic clock, an optical Mach-Zehnder interferometer \cite{lee02}.
The Quantum Rosetta Stone instigated transparent communication between 
the workers in these different areas of research, 
allowing them to adapt the techniques developed in the other areas 
to their own research-greatly advancing all three fields. 
In particular, we focus on the recent developments in 
optical implementations of quantum computing 
and address their utilization in quantum metrology and imaging for 
general audience.

\section{Optical Quantum Computing}

In optical approach to quantum computation, 
qubits are usually represented by the state of polarization of 
a single photon. 
The main difficulty lies in the fact that 
the necessary two-qubit logic gates need a nonlinear interaction 
between the two photons and the efficiency of this nonlinear interaction 
is typically very tiny at the single-photon level. 
This obstacle, however, can be avoided by making corrections 
to the output of the logic devices based on the results 
of single-photon detectors. 
In this section we describe the basic one-qubit operations,
non-deterministic two-qubit gates, 
near-deterministic two-qubit gates,
and certain issues related to their implementations.

\subsection{One-qubit gates}

The one-qubit gates necessary for an arbitrary operation
are any two of x-rotation, y-rotation, and z-rotation,
and a phase gate. The x-rotation, $R_x(\theta)$ is formally written as

\begarr
R_x(\theta) &=& e^{i \sigma_x {\theta \over 2}}
=\cos{\theta \over 2} + i \sigma_x \sin{\theta \over 2}
=\cos{\theta \over 2} + i \pmatrix{ 0 & 1 \cr 1 & 0} \sin{\theta \over 2}
\non \\
&=&
\pmatrix{ \cos{\theta \over 2} & i \sin{\theta \over 2}
 \cr i \sin{\theta \over 2}  & \cos{\theta \over 2} }.
\label{R-x}
\endarr

\no 
Similarly, y-rotation and
z-rotation read

\begarr
R_y(\theta) &=& e^{i \sigma_y {\theta \over 2}}
=\cos{\theta \over 2} + i \sigma_x \sin{\theta \over 2}
=
\cos{\theta \over 2} + i \pmatrix{ 0 & -i \cr i & 0} \sin{\theta \over 2}
\non \\
&=&
\pmatrix{ \cos{\theta \over 2} &  \sin{\theta \over 2}
 \cr - \sin{\theta \over 2}  & \cos{\theta \over 2} },
\\
~\\
R_z(\theta) &=& e^{i \sigma_z {\theta \over 2}}
=\cos{\theta \over 2} + i \sigma_z \sin{\theta \over 2}
=
\cos{\theta \over 2} + i \pmatrix{ 1 & 0 \cr 0 & -1} \sin{\theta \over 2}
\non \\
&=&
\pmatrix{ e^{i \theta \over 2} &  0
 \cr 0  & e^{- {i \theta \over 2}} }.
\endarr

\no
Finally, the phase gate is given by

\begarr
Ph(\theta) 
&=& 
\pmatrix{ e^{i \theta } &  0
 \cr 0  & e^{- i {\theta }} }.
\endarr

Let us suppose now the qubit is encoded in the polarization 
degrees of freedom of a single photon so that the 
logical qubit is given by

\begarr
|0\ra \equiv |H\ra,
\qquad 
|1\ra \equiv |V\ra.
\endarr

\no
{\bf (i) Phase gate}

Then, the z-rotation, for example, can be achieved by a waveplate
and phase shifters.
The wave plate is a 
doubly refracting (birefringent) transparent crystal
where the index of refraction is larger
for the slow axis than the one for the fast axis.
Let us assume that the slow axis is vertical direction and
the fast is the horizontal direction.
After propagating by a distance $d$,
the vertically polarized light acquires a phase shift
of $\exp[i n_1 {2 \pi \over \lambda } d]$, and
similarly, the horizontally polarized light acquires
$\exp[i n_2 {2 \pi \over \lambda } d]$ ($n_1 > n_2$).
The relative phase shift acquired by the vertically polarized light
is then 

\begarr
|V\ra : \Rightarrow e^{i (n_1 - n_2) {\lambda \over 2} d} |V\ra
\equiv e^{i \phi} |V\ra
\endarr

If we choose the thickness of the crystal such that

\begarr
d ={\lambda/4 \over n_1 - n_2} ,
\endarr

\no
then the relative phase shift $\phi =\pi /2$.
This corresponds to the {\em quarter wave plate}
as one can write

\begarr
|H\ra + |V\ra \Rightarrow 
|H\ra + e^{i \pi/2}|V\ra
\endarr

\no
Using the convention for left and right circularly polarized light,

\begarr
\he_- = {1 \over \sqrt{2}} ( \he_1 + i \he_2),
\qquad
\he_+ = {1 \over \sqrt{2}} ( \he_1 - i \he_2)
\endarr

\no
A linearly polarized light ($|H\ra + |V\ra$)
becomes an left circularly polarized light.
In general, an arbitrary phase angle, $\phi$
can be made by choosing the thickness $d$ as

\begarr
d = {\lambda \over n_1 - n_2} { \phi \over 2 \pi}
\endarr

\no
On the other hand,
the overall phase factor also depends on the 
value of $\phi$ such that

\begarr
e^{ i n_2 {2 \pi \over \lambda } d}
&=& e^{i n_2 {2 \pi \over \lambda } {\lambda \over 2 \pi} { \phi \over n_1 - n_2} }
\non
\\
&=&
e^{i {n_2 \over n_1 - n_2} \phi},
\label{overall}
\endarr

\no
{\bf (ii) z-rotation}

Now for the z-rotation, we have

\begarr
R_z(\theta)\big[ |0\ra + |1\ra \big]
&=&
e^{i {\theta \over 2} }|0\ra 
+ e^{ -i {\theta \over 2} }|1\ra
\non
\\
&=&
e^{i {\theta \over 2 }} \big[ |0\ra + e^{-i \theta} |1\ra \big]
\endarr

\no
One can have this operation with a 
wave plate by taking $\phi = -\theta$.
Then, the overall phase factor of Eq.(\ref{overall})
becomes

\begarr
e^{i {n_2 \over n_1 - n_2} \phi}
&=&
e^{-i {n_2 \over n_1 - n_2} \theta}
=\equiv
e^{-i r \theta}
\endarr

\no
where we defined $r={n_2 \over n_1 - n_2}$.
In order to match the overall phase factor of 
the z-rotation ($e^{i {\theta \over 2}}$),
we need to have a phase shifter
such that

\begarr
e^{i \phi'} e^{- i r \theta} = e^{i {\theta \over 2}},
\endarr

\no
yielding

\begarr
\phi' &=& {\theta \over 2} ( 1+ 2 r)
=
{\theta \over 2} {n_1 +n_2 \over n_1 -n_2}
\equiv s \theta
\endarr

\no
Therefore,
the z-rotation is made by a wave plate of $-\theta$ and 
a phase shifter of $s \theta$, where $s=[(n_1 + n_2)/2(n_1 - n_2)]$.
Symbolically, we may write this relation
as 

\begarr
R_z(\theta) :=: {\sf PS}(s\theta) {\sf WP}(-\theta)
\endarr

\no
{\bf (iii) x-rotation}

Now suppose the wave plate we described above is rotated around the propagation
axis by an amount of $\alpha$.
Then the operation by the wave plate can be given as

\begarr
{\sf WP}(\phi, \alpha=0) &=& e^{i r \phi} 
\pmatrix{
1 & 0 \cr 0 & e^{i \phi} },
\non \\
{\sf WP}(\phi, \alpha) &=& 
e^{i r \phi}
\pmatrix{
\cos{\alpha} & -\sin{\alpha} \cr
\sin{\alpha} & \cos{\alpha} }
\pmatrix{
1 & 0 \cr 0 & e^{i \phi} }
\pmatrix{
\cos{\alpha} & \sin{\alpha} \cr
-\sin{\alpha} & \cos{\alpha} }
\non 
\\
&=&
e^{i r \phi}
\pmatrix{
\cos^2{\alpha} + e^{i \phi} \sin^2{\alpha} 
& \cos{\alpha}\sin{\alpha} (1- e^{i \phi}) \cr
\cos{\alpha} \sin{\alpha} (1-e^{i \phi} )
& \sin^2{\alpha}+e^{i \phi} \cos^2{\alpha} }.
\endarr

\no
If we set the rotation angle $\alpha$ equal to
$\pi/4$, we obtain

\begarr
{\sf WP}(\phi, \alpha=\pi/4) &=& 
{e^{i r \phi} \over 2}
\pmatrix{
1 + e^{i \phi}  
& 1- e^{i \phi} \cr
1-e^{i \phi} 
& 1 + e^{i \phi}  }.
\endarr

\no
We can see that the x-rotation 
$R_x(\theta)$ given in Eq.(\ref{R-x}),
can be achieved by
setting $\phi=-\theta$ as

\begarr
{\sf WP}(\phi=-\theta, \alpha=\pi/4) &=& 
e^{-i r \theta}
e^{- i {\theta \over 2}}
\pmatrix{
\cos{\theta \over 2 } & i \sin{\theta \over 2} \cr
i \sin{\theta \over 2} & \cos{\theta \over 2} }.
\endarr

\no
Similarly to the case of $R_z(\theta)$,
we compensate the overall phase factor
by a phase shifter of $e^{i s\theta}$
and then the x-rotation is given by

\begarr
R_x(\theta) &:=:&
{\sf PS}(s \theta)
{\sf WP}(-\theta, \alpha=\pi/4).
\endarr
 
One can then construct the y-rotation by using

\begarr
R_y(\theta) 
&=&
R_z(-{\pi \over 2} )
R_x(\theta)
R_z({\pi \over2 })
\endarr

\no
{\bf (iv) Hadamard gate}

Having the x-rotation, z-rotation,
and the phase gate, it is now sufficient to build any arbitrary one qubit gate.
For example, the Hadamard gate, 
one of the most frequently used one-qubit gates
in the literature, can be build as follows.

\begarr
H &=& {1 \over \sqrt{2}} \pmatrix{
1 & 1 \cr 1 & -1}
\non \\
&=&
Ph(-{\pi \over 2}) R_z(\pi) R_y({\pi \over 2})
\non \\
&=&
Ph(-{\pi \over 2}) R_z(\pi)
R_z(-{\pi \over 2} )
R_x({\pi \over 2})
R_z({\pi \over2 })
\non \\
&=&
Ph(-{\pi \over 2})
R_z({\pi \over 2} )
R_x({\pi \over 2})
R_z({\pi \over2 })
\endarr

\subsection{Two-qubit gates}

A typical two-qubit gate, controlled-NOT (CNOT) is represented as follows:

\begarr
|0\ra_L |0\ra_L &&\rightarrow ~~|0\ra_L |0\ra_L 
\non \\
|0\ra_L |1\ra_L &&\rightarrow ~~|0\ra_L |1\ra_L 
\non \\
|1\ra_L |0\ra_L &&\rightarrow ~~|1\ra_L |1\ra_L 
\non \\
|1\ra_L |1\ra_L &&\rightarrow ~~|1\ra_L |0\ra_L ,
\endarr

\no
where the second qubit (target) flips when
the first qubit (control) has in the logical value 1.
The CNOT gate plays an important role in that any n-qubit gates can be 
decomposed into the CNOT gates and one-qubit gates and thus
form a universal set of gates. 

To physics community, however, the conditional sign flip 
is perhaps more familar. 
The transformation by the conditional sign-flip gate
is written as

\begarr
|0\ra_L |0\ra_L &&\rightarrow ~~|0\ra_L |0\ra_L 
\non \\
|0\ra_L |1\ra_L &&\rightarrow ~~|0\ra_L |1\ra_L 
\non \\
|1\ra_L |0\ra_L &&\rightarrow ~~|1\ra_L |0\ra_L 
\non \\
|1\ra_L |1\ra_L &&\rightarrow ~-|1\ra_L |1\ra_L .
\endarr

\no
The CNOT gate is then simply
constructed by using the conditional sign flip and
two one-qubit gates (e.g., Hadamard on the target, followed by
the conditional sign flip
and another Hadamard on the target).
Note that the CNOT is the controlled $\sigma_x$ operation
whereas the C-Z gate is the controlled $\sigma_z$ operation

The optical Kerr nonlinearity was first considered
for such a conditional sign flip.
The interaction caused by the Kerr nonlinearity can be described by
a Hamiltonian \cite{scully97}
\begarr
{\cal H}_{\rm Kerr} = \hbar \kappa \ad\ha \bd\hb,
\label{kerr}
\endarr
\no
where $\kappa$ is a coupling constant depending on the third-order
nonlinear susceptibility, and 
$\ad$, $\bd$ and $\ha$, $\hb$ are the creation and annihilation 
operators for two optical modes.
One convenient choice of the logical qubit can then be
represented by the two modes containing a single photon,
denoted as
\begarr
|0\ra_L &=& |0\ra_l ~|1\ra_k \non \\
|1\ra_L &=& |1\ra_l ~|0\ra_k ,
\endarr
\no
where 
$l,k$ represent the relevant modes, and
we have used the notation
$|\cdot\ra_L$ for a logical qubit, in order to
distinguish it from the photon-number states $|\cdot\ra_k$.
The conversion between this so-called dual rail qubit and the
polarization qubit can be simply made by polarizing beam splitters.

Now let us assign mode 1,2 for the control
qubit, and 3,4 for the target qubit.
Suppose now only the modes 2,4 are coupled under the 
interaction given by Eq.(\ref{kerr}).
For a given interaction time $\tau$, the transformation
can be written as
\begarr
|0\ra_L |0\ra_L &&\rightarrow ~~|0\ra_L |0\ra_L 
\non \\
|0\ra_L |1\ra_L &&\rightarrow ~~|0\ra_L |1\ra_L 
\non \\
|1\ra_L |0\ra_L &&\rightarrow ~~|1\ra_L |0\ra_L 
\non \\
|1\ra_L |1\ra_L &&\rightarrow e^{i \varphi} |1\ra_L |1\ra_L ,
\endarr
\no
where $\varphi \equiv \kappa n_a n_b \tau$
and 
$n_a =\la \ad\ha \ra, n_b=\la \bd\hb\ra$. 
This operation yields a conditional phase shift.
When $\varphi =\pi$, we have the two two-qubit gate
called the conditional sign-flip gate.
In order to have $\varphi \sim \pi$ at the single-photon level,
however,  a huge third-order nonlinear coupling 
is required \cite{milburn89}.
However, this barrier can be circumvented by quantum teleportation
technique in one hand, and 
effective nonlinearities
produced by projective measurements in the other \cite{klm01}.

\subsection{Nonlinear sign gate}

Together with quantum teleportation
the nonlinear sign (NS) gate serves the
basic element in the architecture of linear optics quantum 
computation.
The NS gate applies to photon number state
consists of zero, one and two photons
as is defined by

\begarr
\alpha|0\ra + \beta |1\ra + \gamma |2\ra
\rightarrow 
\alpha|0\ra + \beta |1\ra - \gamma |2\ra.
\endarr

\begin{figure}
   \begin{center}
   \begin{tabular}{c}
   \includegraphics[height=3cm]{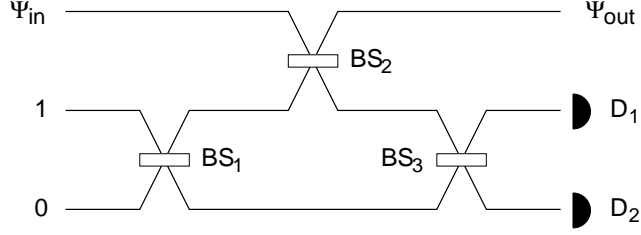}
   \end{tabular}
   \end{center}
   \caption
{\label{ns-1}
Nonlinear sign gate:
The input state of $|\Psi_{\rm in}\ra = \alpha |0\ra
+ \beta |1\ra + \gamma |2\ra$ transforms to
$|\Psi_{\rm out}\ra = \alpha |0\ra
+ \beta |1\ra - \gamma |2\ra$ upon
detecion of one photon at $D_1$ and non-detection
at $D_2$. 
}
\end{figure}

\no
The NS gate can be implemented non-deterministically
by three beam splitters, two photo-detectors,
and one ancilla photon as depicted in Fig.\ref{ns-1}.
Conditioned upon a specific detector outcome, we can
have the desired output state with 
probability of 1/4.
The gate operation succeeds only once in four times
on average.
But, the merit is that we know we it was successful
whenever it was successful.

Now we need to fix the reflection coefficients of the
three beam splitters utilized in the nonlinear sign gate.
Taking $\ha$ and $\hb$ as the input modes,
the output modes can
be found as

\begarr
\pmatrix{ \ha' \cr \hb'}
=
\pmatrix{ r & t \cr t & r}
\pmatrix{\ha \cr \hb}
\endarr

\no
where for lossless beam splitters
the reflection coefficient $r$ and the 
transmission coefficient $t$
need to satisfy
$|r|^2+|t|^2=1$, and
$rt^*+tr^*=0$.
We may define the beam splitter as
\begarr
{\sf BS} (\theta, \phi)
:\equiv:
\pmatrix{
\cos{\theta} & e^{i \phi} \sin{\theta}\cr
e^{i \phi} \sin{\theta} & \cos{\theta} ,
}
\endarr

\no
and the following set of beam splitters 
can to perform the desired operation \cite{ralph01}:

\begarr
{\sf BS}_1 = {\sf BS}_3
&:=:&
\pmatrix{
\sqrt{\eta} &\sqrt{1-\eta} \cr
\sqrt{1-\eta} & -\sqrt{\eta}
},
\qquad
\eta = {1 \over (4 -2\sqrt{2}} \approx 0.854
\\
{\sf BS}_2
&:=:&
\pmatrix{
-\sqrt{\eta_2} &\sqrt{1-\eta_2} \cr
\sqrt{1-\eta_2} & \sqrt{eta_2}
},
\qquad
\eta_2 = (\sqrt{2}-1)^2 \approx 0.172
\endarr

\no
These beam splitters are phase asymmetric,
which can be made by ordinary beam splitters 
with an additional phase shifters so that 
we have

\begarr
{\sf BS}_1 = {\sf BS}_3
&:=:&
\pmatrix{1 & 0 \cr 0 & i}
\pmatrix{
\sqrt{\eta} &-i \sqrt{1-\eta} \cr
-i \sqrt{1-\eta} & \sqrt{\eta}
}
\pmatrix{1 & 0 \cr 0 & i}.
\endarr

\no
This corresponds to two phase shifters of $e^{i \pi/2}$
in the lower path before and after 
the ordinary beam splitters.
Similarly,
for ${\sf BS}_2$ we have

\begarr
{\sf BS}_2
&:=:&
\pmatrix{i & 0 \cr 0 & 1}
\pmatrix{
\sqrt{\eta_2} &-i \sqrt{1-\eta_2} \cr
-i \sqrt{1-\eta_2} & \sqrt{\eta_2}
}
\pmatrix{i & 0 \cr 0 & 1},
\endarr

\no
which 
corresponds to two phase shifters of $e^{i \pi/2}$
in the upper path before and after 
the ordinary beam splitters
with the reflection coefficient
$\eta_2$.

\subsection{Non-deterministic C-Z gate
}

To make the C-Z gate, 
all we need is to change the sign
of the input when the input is corresponding to
$|1\ra_{L} |1\ra_{L}$ state.
Given arbitrary two qubits

\begarr
|Q_1\ra &&= \alpha_0 |0\ra_{L} + \alpha_1 |1\ra_{L}
= \alpha_0 |0\ra_1 |1\ra_2 + \alpha_1 |1\ra_1 |0\ra_2 ,
\non \\
|Q_2\ra &&= \alpha_0^\prime |0\ra_{L} + \alpha_1^\prime
|1\ra_{L}
=
\alpha_0^\prime |0\ra_3 |0\ra_4 + \alpha_1^\prime |1\ra_3 |0\ra_4 ,
\endarr

\no
the transformation by applying a C-Z gate, can 
be written as

\begarr
|Q_1\ra |Q_2\ra
&&\Rightarrow
\alpha_0 \alpha_0^\prime |0\ra_{L} |0\ra_L 
+ \alpha_0 \alpha_1^\prime |0\ra_L |1\ra_{L}
+ \alpha_1 \alpha_0^\prime |1\ra_L |0\ra_L 
- \alpha_1 \alpha_1^\prime |1\ra_L |1\ra_L 
\non \\
&&=
\alpha_0 \alpha_0^\prime |0,1,0,1\ra
+ \alpha_0 \alpha_1^\prime |0,1,1,0\ra
+ \alpha_1 \alpha_0^\prime |1,0,0,1\ra
- \alpha_1 \alpha_1^\prime |1,0,1,0\ra  .
\endarr

\begin{figure}
   \begin{center}
   \begin{tabular}{c}
   \includegraphics[height=3.5cm]{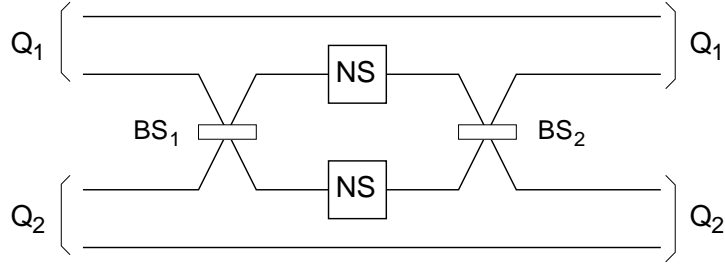}
   \end{tabular}
   \end{center}
   \caption
{\label{cz-1}
C-Z gate:
The implementation of C-Z gate is made by a combination
of the nonlinear sign gate and the physics of
Hong-Ou-Mandel (HOM) interferometer.
}
\end{figure}

\no
where
the modes 1, 2 are designated for the control qubit,
and 3, 4 are for the target qubit, and
the number state representation was given in this
order (see Fig.\ref{cz-1}).  
We can immediately see that when there is a sign change
only when there is one photon in mode 1
and one photon in mode 3.

Here the two additional 50/50 beam splitters (shown in the figure)
need to be mutually conjugate
given by

\begarr
{\sf BS}_1 &=& {\sf BS}(\theta =\pi/2, \phi=-\pi/2)
:=:
{1 \over \sqrt{2}}
\pmatrix{1 & -i \cr -i & 1}
\\
{\sf BS}_2 &=& {\sf BS}(\theta =\pi/2, \phi=\pi/2)
:=:
{1 \over \sqrt{2}}
\pmatrix{1 & i \cr i & 1}
\endarr

\no
To summarize, we need
six asymmetric beam splitters (or six ordinary symmetric beams plitters and 
six phase shifters), four single-photon detectors,
and two mutually conjugate 50/50 beam splitters for the C-Z gate.
In addition 
four polarization beam splitters
in order to convert
the polarization qubits to the dual-rail qubits
back and forth.
This is a non-deterministics gate in that the operation is successful
only one in sixteen times on average .
The probability of success then can be boosted by using gate-teleportation 
technique and sufficient number of ancilla photons.
It has been also demonstrated that such a nondeterministic two-qubit 
gate can be made for qubits defined by the 
polarization degree of freedom \cite{pittman01,pittman02}.
Reliable single-photon sources and the ability to  
discriminate the number of incoming photons play the essential roles in the
realization of such nonlinear quantum gates in optical quantum computing.

\section{Quantum Metrology and Imaging}

\begin{figure}
   \begin{center}
   \begin{tabular}{c}
   \includegraphics[height=3cm]{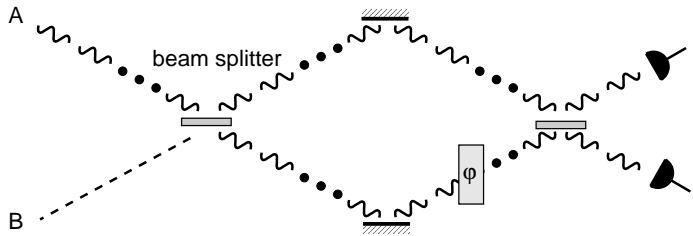}
   \end{tabular}
   \end{center}
   \caption
   {\label{mz-1}
A typical Mach-Zehnder interferometer
where the input light is incident on input port A while only
vacuum comes into input port B.
}
\end{figure}

In a Mach-Zehnder interferometer, the input light field is divided
into two different paths by a beam splitter, and recombined by another
beam splitter. The phase difference between the two paths is then
measured by balanced detection of the two output modes (see Fig.~\ref{mz-1}).
We may think of the upper and lower paths as the two states
in which a single light quanta can occupy.
With a coherent laser field as the input the phase
sensitivity is given by the shot noise limit $1/\sqrt{N}$, where
$N$ is the average number of photons passing though the interferometer
during measurement time. When the number of photons is exactly known
(i.e., the input is a Fock state $|N\rangle$), the phase sensitivity
is still given by $1/\sqrt{N}$, indicating that the photon
counting noise does not originate from the intensity fluctuations of
the input beam, but rather from the Poissonian ``sorting noise'' of
the beam splitter.

There have been various proposals for achieving Heisenberg-limited
sensitivity, depending on the input states and the detection schemes.
In particular, the Yurke state approach has the same measurement scheme as
the conventional Mach-Zehnder interferometer;
a direct detection of the difference current \cite{yurke86a,yuen86,yurke86b,dowling98}.
It is, however, not easy to generate the desired correlation
in the input state.
On the other hand, the dual Fock-state
approach finds a rather simple input state, but
requires a complicated data processing methods \cite{holland93,kim98}.
However, by a simple utilization of the projective measurements 
with linear optical devices,
it is possible to generate a desired
correlation in the Yurke state 
directly from the dual Fock state.

\begin{figure}[ht]
   \begin{center}
   \begin{tabular}{c}
   \includegraphics[height=3.5cm]{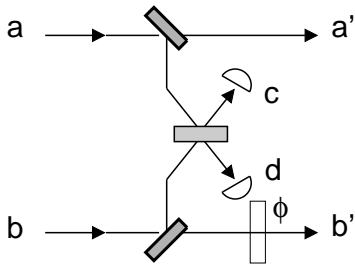}
   \end{tabular}
   \end{center}
   \caption{
   \label{y-1}
   A Yurke-type quantum correlation between the two modes can be
produced with a dual Fock-state.
Suppose we post-select the outcome, conditioned upon
only one photon detection by either one of the two detectors.
Due to the 50:50 beam splitter in the midway, it is not possible to tell
whether mode $a$ or $b$ lost one photon.
}
\end{figure}

Let us consider a linear optical setup depicted 
in Fig.\ref{y-1}.
For a given dual Fock-state input $|N,N\ra_{ab}$, 
the output state conditioned on, for example, a
two-fold coincident count is given by
\begarr
 {\textstyle \frac{1}{\sqrt{2}}} \left[ |N,N-2\ra+ |N-2,N\ra \right] .
 \label{dowling}
\endarr
\no
It is not difficult to see that the condition of the coincident
detection yields either one of the two modes before the beam 
splitter must contain two photons while the other modes
contains no photon. 
This is an inverse-HOM situation where 
one photon at each mode cannot contribute 
to the coincident detection.
Consequently, 
the coincident detection results in a situation
where the main modes
$a$ and $b$ can only lose two photons or not at all.

Here the probability success of this event can be optimized by choosing
the reflection coefficient of the first beam splitters.
For the reflection coefficient of $|r|^2 = {1 / N}$,
its asymptotic value is found as $1/(2e^2)$, independent of $N$.
Furthermore, using a stack of such devices 
with appropriate phase shifters,
we have developed a method for the generation of 
maximally path-entangled states of the form 
$|N,0\ra_{ab}+|0,N\ra_{ab}$, the so-called NOON state,
with an arbitrary number of photons \cite{kok02a}.

Quantum correlations can also be applied to optical lithography. 
In recent work it has been shown that the Rayleigh diffraction limit
in optical lithography can be circumvented by the use of
path-entangled photon number states \cite{boto00}. The desired
$N$-photon path-entangled state, for $N$-fold resolution enhancement,
is the NOON states.

Consider the simple case of a two-photon Fock state $|1\ra_A |1\ra_B$,
which is a natural component of a spontaneous parametric
down-conversion event. After passing through a 50/50 beam splitter,
it becomes an entangled number state of the form $|2\ra_A |0\ra_B +
|0\ra_A |2\ra_B$. Interference suppresses the probability amplitude 
of $|1\ra_A |1\ra_B$. According to quantum mechanics, it is not
possible to tell whether both photons took path $A$ or $B$ after the
beam splitter. 

When parametrizing the position $x$ on the surface by $\varphi=\pi
x/\lambda$, the deposition rate of the two photons onto the substrate
becomes $1+\cos 2\varphi$, which has twice better resolution
$\lambda/8$ than that of single-photon absorption, $1 + \cos\varphi$,
or that of uncorrelated two-photon absorption, $(1 +\cos\varphi)^2$.
For the NOON state we obtain
the deposition rate $1+\cos N\varphi$, corresponding to a resolution
enhancement of $\lambda/(4N)$. 

It is well known that the two-photon NOON state can be 
generated using a Hong-Ou-Mandel (HOM)
interferometer \cite{hom87} and two single-photon input states.
A 50/50 beam splitter, however, is not sufficient for producing
path-entangled states with a photon number larger than two.

\vspace{1cm}

\section{Universal NOON State Generation}

\begin{figure}[ht]
   \begin{center}
   \begin{tabular}{c}
\hspace*{2cm}
\includegraphics[height=4.5cm]{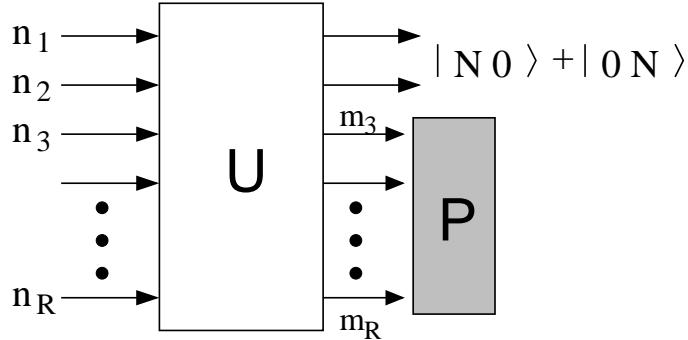}
  \end{tabular}
   \end{center}
\caption{
\label{noon}
NOON state generating machine. 
$R$ input modes, fed with $n_{i}, i=1,...,R$ photons each, 
are transformed into a linear superposition of the output modes 
by a unitary operation $U$. 
A successive projective measurement $P$ of $m_{j}, j=3,..,R$ photons 
in $R-2$ output modes leads to a collapse of the wave function 
into a NOON state in the unmeasured output modes 1 and 2.}
\end{figure}

As we have mentioned earlier, one of the possible ways to generate 
a two-mode entangled many-photon state, 
such as a NOON state, is to make use of an optical nonlinearity. 
However, due to a relatively small average number of photons involved 
an overall effect of the nonlinearity is extremely tiny and 
practically one cannot make use of it. 
An alternative way to enable an effective photon-photon interaction is 
to use ancilla modes and projective measurements. 
In this case a {\em universal NOON state generating machine} 
can be built consisting out of two main blocks Fig.\ref{noon}.


The first part is a $R$-port unitary device producing 
a linear superposition of the $R-2$ input ancilla modes 
and two target input modes. 
The second part is a non-unitary projection device 
performing a photodetection of $m_{j}, j=2,...,R$ photons 
in each of the output ancilla modes. 
In this way, an interplay between the number of input photons, 
parameters of the $R$-port linear device and 
the number of detected output ancilla photons 
determines a state of two output target modes.

In general there is a limitation on a possible size of 
an output NOON state in such a generating machine.  
Here we formulate the following ``no-go theorem'': 
Suppose we have an $R$-port NOON state generating machine as depicted in Fig.\ref{noon}. 
The NOON state can be generated in the two output modes 
with at most $R$ photons. 
In other words, in order to generate a NOON state $|\psi\rangle=\frac{1}{\sqrt{2}}(|N,0\rangle+|0,N\rangle)$ 
one requires at least  $N$ input modes. 
It means that the number of the resources needed 
to build a high-NOON state (a NOON state with high N)
grows polynomially in $N$. However, the success probability drops down exponentially.

As an example illustrating our statements 
we consider a NOON state generator with four input modes, 
i.e., the number of ancilla modes is 2. 
According to the theorem formulated above, 
we can generate at most a four-photon NOON state 
$|\psi\rangle=\frac{1}{\sqrt{2}}(|4,0\rangle + |0,4\rangle)$. 
This can be done in a number of different ways 
with different success probabilities. 
Depending on the availability of the resources and photodetection preferences 
we can construct different realizations of the NOON state generator. 

~

\begin{figure}[hc]
   \begin{center}
   \begin{tabular}{c}
\includegraphics[height=6.5cm]{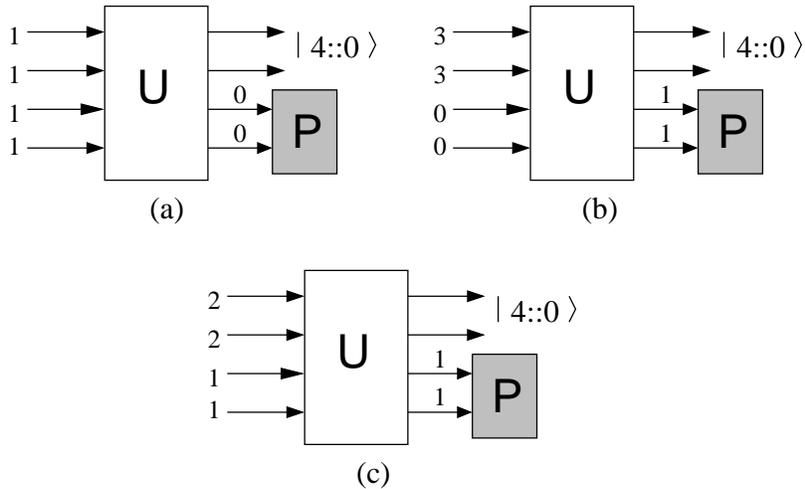}
  \end{tabular}
   \end{center}
\caption{
\label{noon-4}
Possible allowed configurations for the four-photon NOON state generation.
$|4::0\ra$ denotes $\frac{1}{\sqrt{2}}(|4,0\ra + |0,4\ra)$\cite{kok02a}.
}
\end{figure}

If one wants, for example, to rely on a vacuum detection
(or, in other words, no-photons detection) in both output ancilla modes, 
then one can realize a scheme depicted on Fig.\ref{noon-4}(a). 
In this case each input mode is fed with only one photon 
and the four-photon NOON state is generated upon a vacuum measurement 
in both output ancilla modes\cite{fiurasek02}. 
Alternatively, a scheme based on a coincidence detection 
in the output ancilla modes is presented on Fig.\ref{noon-4}(b).
Here, the input ancilla modes contain zero photons \cite{lee02a}.  
A scheme based on a coincidence detection and non-zero ancilla input 
is also possible and depicted on Fig.\ref{noon-4}(c). 
The variety of possible realizations of a given NOON state 
sets up a question about the most optimal configuration of a NOON state generator.

\newpage


\end{document}